\documentclass[12pt]{iopart}

\usepackage{iopams}
\usepackage{epsfig}
\usepackage{graphicx}
\usepackage{color}
\usepackage{bm}
\usepackage[bookmarks, pdfstartview=FitH]{hyperref}
\usepackage{hypernat}

\begin{document}

\title{Two-qubit gate operations in superconducting circuits with strong coupling and weak anharmonicity}

\author{Xin-You L\"{u}$^{1,3}$, S. Ashhab$^{1,2}$, Wei Cui$^{1}$, Rebing Wu$^{1,4}$, Franco Nori$^{1,2}$}
\address{$^1$Advanced Science Institute, RIKEN, Wako-shi, Saitama 351-0198, Japan\\$^2$Physics Department, The University of Michigan, Ann Arbor, Michigan 48109-1040, USA
\\$^3$School of Physics, Ludong University, Yantai 264025, P. R. China
\\$^4$Department of Automation, Center for Quantum information Science and Technology, Tsinghua University, Beijing, 100084, P.R. China}

\begin{abstract}
We investigate theoretically the implementation of two-qubit gates
in a system of two coupled superconducting qubits. In particular, we
analyze two-qubit gate operations under the condition that the
coupling strength is comparable to or even larger than the
anharmonicity of the qubits. By numerically solving the
time-dependent Schr\"{o}dinger equation under the assumption of
negligible decoherence, we obtain the dependence of the two-qubit
gate fidelity on the system parameters in the case of direct and
indirect qubit-qubit coupling. Our numerical results can be used to
identify the ``safe'' parameter regime for experimentally
implementing two-qubit gates with high fidelity in these systems.
\end{abstract}
\pacs{03.67.-a; 42.50.Pq; 85.25.-j}

 \maketitle

\section{Introduction}
Superconducting (SC) circuits based on Josephson junctions are
promising candidates for the realization of scalable quantum
computing on a solid-sate platform, due to their design flexibility,
large-scale integration and controllability (see the reviews in
Refs.\,\cite{1,2,3,4,5,6,7}). SC qubits, include the charge
\cite{8}, flux \cite{9}, and phase qubits \cite{10,10.5} as well as
their variants, capacitively shunted flux qubits \cite{11} and
capacitively shunted charge qubits (transmon) \cite{11.5}. The phase
qubit, the capacitively shunted flux qubit and the transmon qubit
are relatively insensitive to charge noise and can be operated over
a wide range of parameters. Single-qubit gates \cite{12}, two-qubits
gates \cite{13,14} and simple quantum algorithms \cite{15} with
these types of qubits have been demonstrated experimentally in
recent years. However, comparing with the flux qubits, the common
disadvantage of these types of qubits is their weakly-anharmonic
energy level structure, i.e., the detuning between adjacent
transition frequencies is very small.

Generally, the influence of the small anharmonicity (denoted by
$\Delta$) on quantum gate operations can be neglected when the
qubit-field or qubit-qubit coupling strength is very small compared
with $\Delta$. However, for the practical application of quantum
computation, one wants to maximize the number of quantum gate
operations with a given coherence time. In other words, we must
implement quantum operations as fast as possible, which requires a
strong qubit-qubit or qubit-field coupling to be employed during the
single- and two-qubit gate operations \cite{15.5}. The anharmonicity
of SC qubits will influence the quality of quantum gates more and
more with increasing coupling strength. Recently, there have been a
number of theoretical studies analyzing the effects of weak
anharmonicity of SC qubits on the operation of single-qubit gates
and several optimization strategies have been proposed based on
varying driving pulse shapes and sequences \cite{16,17,18,19,20}.
Similar to single-qubit gates, the weak anharmonicity of SC qubits
will also influence the implementation of two-qubit gates. Then two
questions arise naturally: (1) how much the weak anharmonicity of
the qubits influence the implementation of two-qubit gates in a
system of coupled SC qubits? (2) how strong can the coupling be
while allowing a high two-qubit gate fidelity? In other words, how
fast can two-qubit gates with high fidelity be implemented, given
the weak anharmonicity of SC qubits?

Motivated by the above questions, in this paper we study the
implementation of two-qubit gates with superconducting systems in
the strong coupling regime. First, we introduce some possible
methods for implementing two-qubit gates and qualitatively discuss
the effect of strong coupling (section II). Then, in section III, we
numerically simulate the influence of the coupling strength and
anharmonicity on the fidelities of two-qubit gates in different
superconducting systems, and show that the ``safe'' parameter regime
for implementing two-qubit gates with high fidelity can be
identified, which is useful for guiding experimental efforts based
on superconducting qubits. Finally, we conclude with a brief summary
in section IV.

\begin{figure}
\centering
\includegraphics[width=10cm]{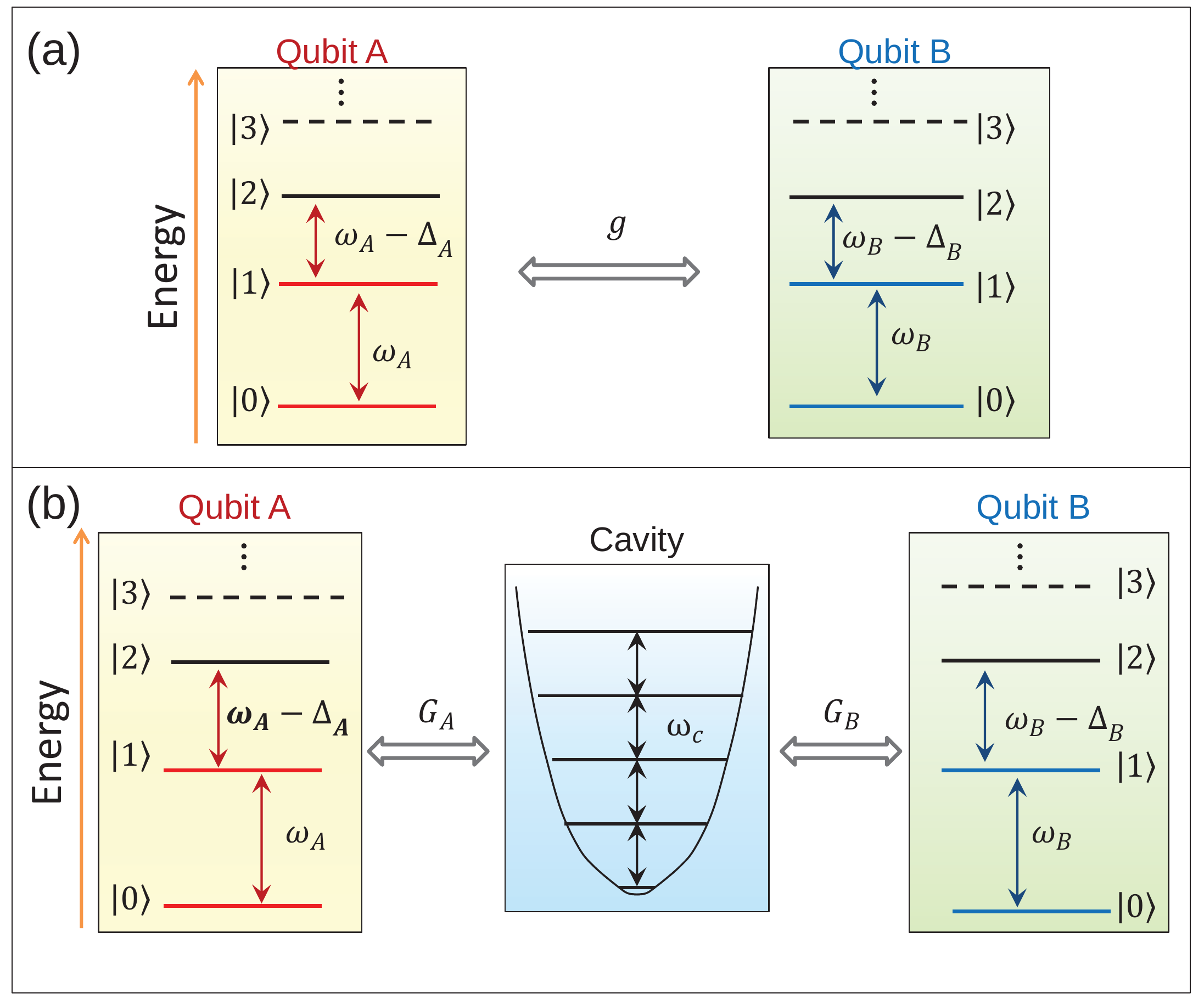} \caption{(Color
online) System with direct (a) and indirect (b) qubit-qubit
coupling. Here, $g$, $G_{j}$ and $\Delta_{j}$ $(j=A,B)$ are the
qubit-qubit, qubit-cavity coupling strength and anharmonicity,
respectively.}
\end{figure}
\section{Model and qualitative discussion}
As shown in Fig.\,1, as model systems we consider two directly (a)
or indirectly (b) coupled SC qubits with weakly-anharmonic
multilevel structure (such as transmon or phase qubits). Here it
should be pointed out the flux qubits have a strong anharmonicity,
and the problem discussed in this paper is not a serious limitation.
The two lowest levels $\{|0\rangle_{j}$, $|1\rangle_{j}\}$,
separated in energy by $\hbar \omega_{j}$ ($j=A,B$), are the
computational basis, and the $n$th ($n\geq2$) higher levels are
different from $n\hbar \omega_{j}$ by $\hbar \epsilon^{j}_{n}$. Here
$\epsilon^{j}_{n}$ has the standard nonlinear oscillator form
$\epsilon^{j}_{n}=\Delta_{j}(n-1)n/2$ \cite{21} and $\Delta_{j}$ is
the anharmonicity of the qubit, and it is positive in our paper.

In the case of direct qubit-qubit coupling, two qubits are directly
(capacitively) coupled, while they are dispersively coupled to a
common transmission line resonator in the case of indirect
qubit-qubit coupling. The Hamiltonian of these two types of coupled
system is given by ($\hbar=1$) \cite{22,22.5,23,24,25,26,27,28}
\numparts
\begin{eqnarray}
H^{direct}&=&\sum^{N-1}_{n=1}\left[\left(n\omega_{A}-\epsilon^{A}_{n}\right)|n\rangle_{A}\langle
n|+\left(n\omega_{B}-\epsilon^{B}_{n}\right)|n\rangle_{B}\langle
n|\right]+gJ^{x}_{A}\otimes J^{x}_{B},
\\
H^{indirect}&=&\omega_{c}a^{\dag}a+\sum_{j=A,B}\left[\sum^{N-1}_{n=1}\left(n\omega_{j}-\epsilon^{j}_{n}\right)|n\rangle_{j}\langle
n|+G_{j}(a+a^{\dag})J^{x}_{j}\right],
\\
J^{x}_{A}&=&\sum^{N-1}_{n=1}\eta^{A}_{n-1,n}\sigma^{Ax}_{n-1,n},\;\;\;J^{x}_{B}=\sum^{N-1}_{n=1}\eta^{B}_{n-1,n}\sigma^{Bx}_{n-1,n},
\end{eqnarray}
\endnumparts
where $H^{d}$ and $H^{id}$ denote the Hamiltonian for the system
with direct and indirect qubit-qubit coupling, $N$ is the number of
levels in each SC qubit, $\eta^{j}_{n-1,n}=\sqrt{n}$ is the
level-dependent coupling matrix element, and
$\sigma^{jx}_{n-1,n}=|n-1\rangle_{j}\langle
n|+|n-1\rangle_{j}\langle n|$ is the effective Pauli spin operators
for levels $|n-1\rangle$ and $|n\rangle$. Also, $\omega_{c}$ is the
frequency of the quantized cavity mode;  $g$ and $G_{j}$ denote the
qubit-qubit and qubit-cavity coupling strength.

In order to qualitatively analyze the implementation and fidelity of
two-qubit gates, we assume that each qubit has three levels. Then,
the Hamiltonian of direct qubit-qubit coupled system ($H^{direct}$),
under the rotation-wave approximation (RWA), can be reduced to
\begin{eqnarray}
\fl \;\;\;\;\;\;
H^{direct}_{I}=\sum_{j=A,B}\left[\omega_{j}|1\rangle_{j}\langle
1|+\left(2\omega_{j}-\Delta_{j}\right)|2\rangle_{j}\langle
2|\right]\nonumber
\\
+g[|01\rangle\langle10|+\sqrt{2}|02\rangle\langle11|+\sqrt{2}|20\rangle\langle11|+2|12\rangle\langle21|+h.c.],
\end{eqnarray}
where $|mn\rangle$ denotes $|m\rangle_{A}|n\rangle_{B}$.

For the system with indirect qubit-qubit coupling, under the
dispersive qubit-cavity-coupling condition, i.e.,
$\mid\delta_{j}\mid=\mid\omega_{j}-\omega_{c}\mid\gg G_{j}$
$(j=A,B)$, the qubits will exchange energy by virtual photon
processes. Then we can obtain the Hamiltonian of the effective
qubit-qubit interaction by a Fr\"{o}hlich transformation
\cite{29,30,31,32},
\begin{eqnarray}
\fl H^{indirect}_{{\rm
eff},1}=\exp(-S)H^{id}\exp(S)\nonumber\\
\fl
\;\;\;\;\;\;\;\;\;\approx\sum_{j=A,B}\left\{\left[\left(\omega_{j}+\frac{G^{2}}{\delta_{j}}\right)|1\rangle_{j}\langle1|+\left(2\omega_{j}-\Delta_{j}+\frac{2G^{2}}{\delta_{j}-\Delta_{j}}\right)|2\rangle_{j}\langle2|
+\frac{G^{2}}{2\delta_{j}}a^{\dag}a\left(|1\rangle_{j}\langle1|-|0\rangle_{j}\langle0|\right)\right.\right.\nonumber\\
\fl
\;\;\;\;\;\;\;\;\;\left.\left.+\frac{G^{2}}{\delta_{j}-\Delta_{j}}a^{\dag}a\left(|2\rangle_{j}\langle2|-|1\rangle_{j}\langle1|\right)\right]+\left[\frac{\sqrt{2}G^{2}}{2}\left(\frac{1}{\delta_{j}-
\Delta_{j}}-\frac{1}{\delta_{j}}\right)a^{2}|2\rangle_{j}\langle0|\right.\right.\nonumber\\
\fl
\;\;\;\;\;\;\;\;\;\left.\left.+\frac{G^{2}}{2}\left(\frac{1}{\delta_{A}}+\frac{1}{\delta_{B}}\right)|01\rangle\langle10|+\frac{\sqrt{2}G^{2}}{2}\left(\frac{1}{\delta_{B}-\Delta_{B}}+\frac{1}{\delta_{A}}\right)|02\rangle\langle11|
\right.\right.\nonumber\\
\fl
\;\;\;\;\;\;\;\;\;\left.\left.+\frac{\sqrt{2}G^{2}}{2}\left(\frac{1}{\delta_{A}-\Delta_{A}}+\frac{1}{\delta_{B}}\right)|20\rangle\langle11|+G^{2}\left(\frac{1}{\delta_{A}-\Delta_{A}}+\frac{1}{\delta_{B}-\Delta_{B}}\right)|12\rangle\langle21|+h.c.\right]\right\},
\end{eqnarray}
where
\begin{eqnarray}
S=\sum_{j=A,B}\left[\frac{G}{\delta_{j}}a^{\dag}|0\rangle_{j}\langle1|+\frac{\sqrt{2}G}{\delta_{A}-\Delta_{A}}a^{\dag}|1\rangle_{j}\langle2|-h.c.\right].
\end{eqnarray}
Here, we have assumed that $G_{A}=G_{B}=G$.
\begin{figure}[here]
\centering
\includegraphics[width=11cm]{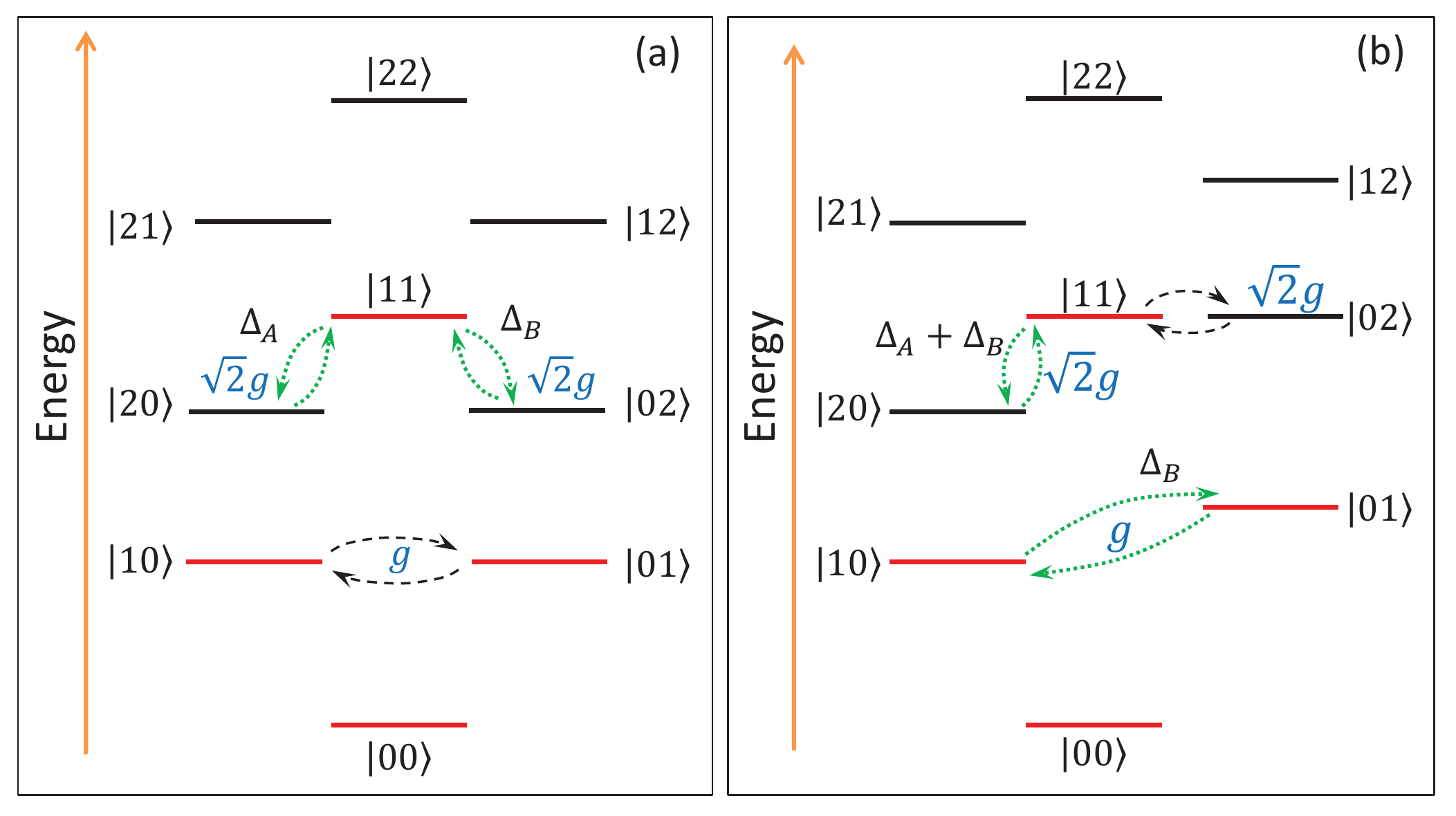}
\caption{(Color online) The energy-level diagram of two-qubit
product states for the iSWAP gate (a), and the controlled-Z gate (b)
in the system with direct qubit-qubit coupling. Red levels denote
the states in the computational basis. The black dashed arrows are
the resonant transitions used for realizing the two-qubit gates and
the green dotted arrows are the main $undesired$ transitions, which
adversely affect the implementation of two-qubit gates. The
couplings $g$ and $\sqrt{2}g$ are indicated in blue, while the
detuning between levels is indicated in black. This figure also
applies the system with indirect qubit-qubit coupling when the
corresponding couplings is replaced by $g_{{\rm eff},m}$
$(m=1,2,3,4)$.}
\end{figure}

The terms proportional to $G^{2}$ in the first four terms of
equation (3) represent level shifts, and the fifth term describes
two-photon processes. Under the dispersive qubit-cavity-coupling
condition, the cavity mode is only virtually excited during the gate
operation, and therefore the third, fourth, and fifth terms of
equation (3) vanish. Then, the Hamiltonian (3) can be simplified
further as \cite{32.1,32.2,32.3,32.4,32.5}
\begin{eqnarray}
\fl H^{indirect}_{{\rm
eff},2}=\sum_{j=A,B}\left[\tilde{\omega}_{j1}|1\rangle_{j}\langle
1|+\left(\tilde{\omega}_{j2}-\Delta_{j}\right)|2\rangle_{j}\langle
2|\right]\nonumber
\\
\fl \;\;\;\;\;\;\;\;\;\;\;\;\;+\left[\sqrt{2}g_{{\rm
eff},1}|02\rangle\langle11|+\sqrt{2}g_{{\rm
eff},2}|20\rangle\langle11|+g_{{\rm
eff},3}|01\rangle\langle10|+2g_{{\rm
eff},4}|12\rangle\langle21|+h.c.\right].
\end{eqnarray}
where
\numparts
\begin{eqnarray}
\tilde{\omega}_{j1}=\omega_{j}+\frac{G^{2}}{\delta_{j}},
\\
\tilde{\omega}_{j2}=2\omega_{j}+\frac{2G^{2}}{\delta_{j}-\Delta_{j}},
\end{eqnarray}
\begin{eqnarray}
g_{{\rm
eff},1}&=&\frac{G^{2}}{2}\left(\frac{1}{\delta_{B}-\Delta_{B}}+\frac{1}{\delta_{A}}\right),
\\
g_{{\rm
eff},2}&=&\frac{G^{2}}{2}\left(\frac{1}{\delta_{A}-\Delta_{A}}+\frac{1}{\delta_{B}}\right),
\\
g_{{\rm
eff},3}&=&\frac{G^{2}}{2}\left(\frac{1}{\delta_{A}}+\frac{1}{\delta_{B}}\right),
\\
g_{{\rm
eff},4}&=&\frac{G^{2}}{2}\left(\frac{1}{\delta_{A}-\Delta_{A}}+\frac{1}{\delta_{B}-\Delta_{B}}\right).
\end{eqnarray}
\endnumparts
Now, we obtain an effective interaction Hamiltonian similar to the
Hamiltonian (2) in the system with direct qubit-qubit coupling.

From the Hamiltonians (2) and (5), it is easily seen that various
two-qubit gates can be realized by appropriately adjusting the qubit
frequencies ($\omega_{A}$, $\omega_{B}$) both in the system with
direct and indirect qubit-qubit coupling. For example, by setting
$\omega_{A}=\omega_{B}$ ($\omega_{B}=\omega_{A}+\Delta_{B}$), the
resonant transition between state $|01\rangle$ and $|10\rangle$
($|11\rangle$ and $|02\rangle$) can be obtained as shown in Fig.\,2.
Then the two-qubit iSWAP \cite{13} (CZ \cite{14,15}) gate can be
realized after an interaction time $gt_{g}=\pi/2$ or $g_{{\rm
eff},3}t_{g}=\pi/2$ ($\sqrt{2}gt=\pi$ or $\sqrt{2}g_{{\rm
eff},1}t=\pi$). Here it should be pointed out that some undesired
transitions [see the (green) dotted arrows in Fig.\,2] have been
neglected in the weak-coupling regime $g\ll|\Delta_{j}|$ or $g_{{\rm
eff},m}\ll|\Delta_{j}|$ $(m=1-4; j=A,B)$. However with increasing
coupling strength $g$ or $g_{{\rm eff},m}$, the average amplitude
$g/|\Delta_{j}|$ or $g_{{\rm eff},m}/|\Delta_{j}|$ of undesired
transitions will become larger and larger, which can not be
neglected again and will reduce the fidelity of the two-qubit gate.
So, the relative value of the coupling strength $g$ or $g_{{\rm
eff},m}$ and the anharmonicity $\Delta_{j}$ is an important
parameter for the quality of the two-qubit gate. In the two-qubit
gate scheme based on SC qubits, a very strong qubit-qubit or
qubit-cavity coupling strength cannot be employed due to the weak
anharmonicity of the qubits, if one wants to obtain a high fidelity.
How strong the coupling can be, while allowing high two-qubit-gate
fidelities, will be analyzed in detail in the next section.

\section{Numerical results}
In this section, we will numerically calculate the fidelity of
two-qubit gates in the circuits with either direct or indirect
qubit-qubit coupling. Importantly, the present numerical results can
help identify the safe parameter regime for implementing two-qubit
gates with high fidelity. Here, we neglect the noise and decoherence
of system in order to show explicitly the influence of coupling
strength and anharmonicity on the fidelity of two-qubit gates. Here,
it should also be pointed out that the single-qubit gates are
performed using microwave pulses (with frequencies of a few of GHz),
while the frequency tuning for the two-qubit gates are implemented
using trapezoidal pulses.

Here, the fidelity of a two-qubit gate is defined as the Euclidean
distance between the target $U_{T}$ and the actual evolution
$U(t_{g})$ \cite{19},
\begin{eqnarray}
F=1-\frac{1}{16}\|U_{T}-P^{\dag}U(t_{g})P \|^{2}_{2},
\end{eqnarray}
where $U(t)$ is the usual time evolution operator obeying the
Schr\"{o}dinger equation $\dot{U}(t)=-\frac{i}{\hbar}H(t)U(t)$ in
the full space of the quantum system. Here $\|X\|_{2}^{2}={\rm
tr}(X^{\dag}X)$ where $X$ is an arbitrary operator. $P$ is the
projection operator on the two-qubit computational $\{|00\rangle,
|01\rangle, |10\rangle, |11\rangle\}$;
$$U_{T}=|00\rangle\langle00|-i|01\rangle\langle10|-i|10\rangle\langle01|+|11\rangle\langle11|$$
corresponds to the two-qubit iSWAP gate, and
$$U_{T}=|00\rangle\langle00|+|01\rangle\langle01|+|10\rangle\langle10|-|11\rangle\langle11|$$
corresponds to the two-qubit CZ gate. Here it should be pointed out
that single-qubit rotations and an overall phase factor
$U^{A}_{z}=e^{i\theta_{A}\sigma^{A}_{z}}$,
$U^{B}_{z}=e^{i\theta_{B}\sigma^{B}_{z}}$, $U_{I}=e^{i\theta I}$ are
used in the numerical calculations in order to eliminate any extra
phase factors; $I$ is the unit matrix and
$$\sigma^{A}_{z}=|00\rangle\langle00|+|01\rangle\langle01|-|10\rangle\langle10|-|11\rangle\langle11|,$$
$$\sigma^{B}_{z}=|00\rangle\langle00|-|01\rangle\langle01|+|10\rangle\langle10|-|11\rangle\langle11|.$$
Specifically, in our numerical calculations, we replace the unitary
operation $U(t_{g})$ in Eq.\,(7) by
$U'(t_{g})=U_{I}U^{B}_{z}U^{A}_{z}U(t_{g})$ and choose $\theta_{A}$,
$\theta_{B}$ and $\theta$ that maximize the fidelity.

We also note here that in our numerical calculations we do not use
the RWA. But, there is almost no difference between these results
shown below and the numerical results with the RWA (not shown in
this paper). The reason is that the parameter regime that we
consider does not reach the ultrastrong coupling regime and thus the
RWA is valid here. Very recently, the influence of the
counter-rotating terms in the Hamiltonian on the two-qubit gates in
the ultrastrong coupling regime has been studied in a related system
\cite{33}. Also, the effect of counter-rotating terms were  studied
in \cite{33.5}.

\subsection{System with direct qubit-qubit coupling}

\begin{figure}
\centering
\includegraphics[width=0.76\textwidth]{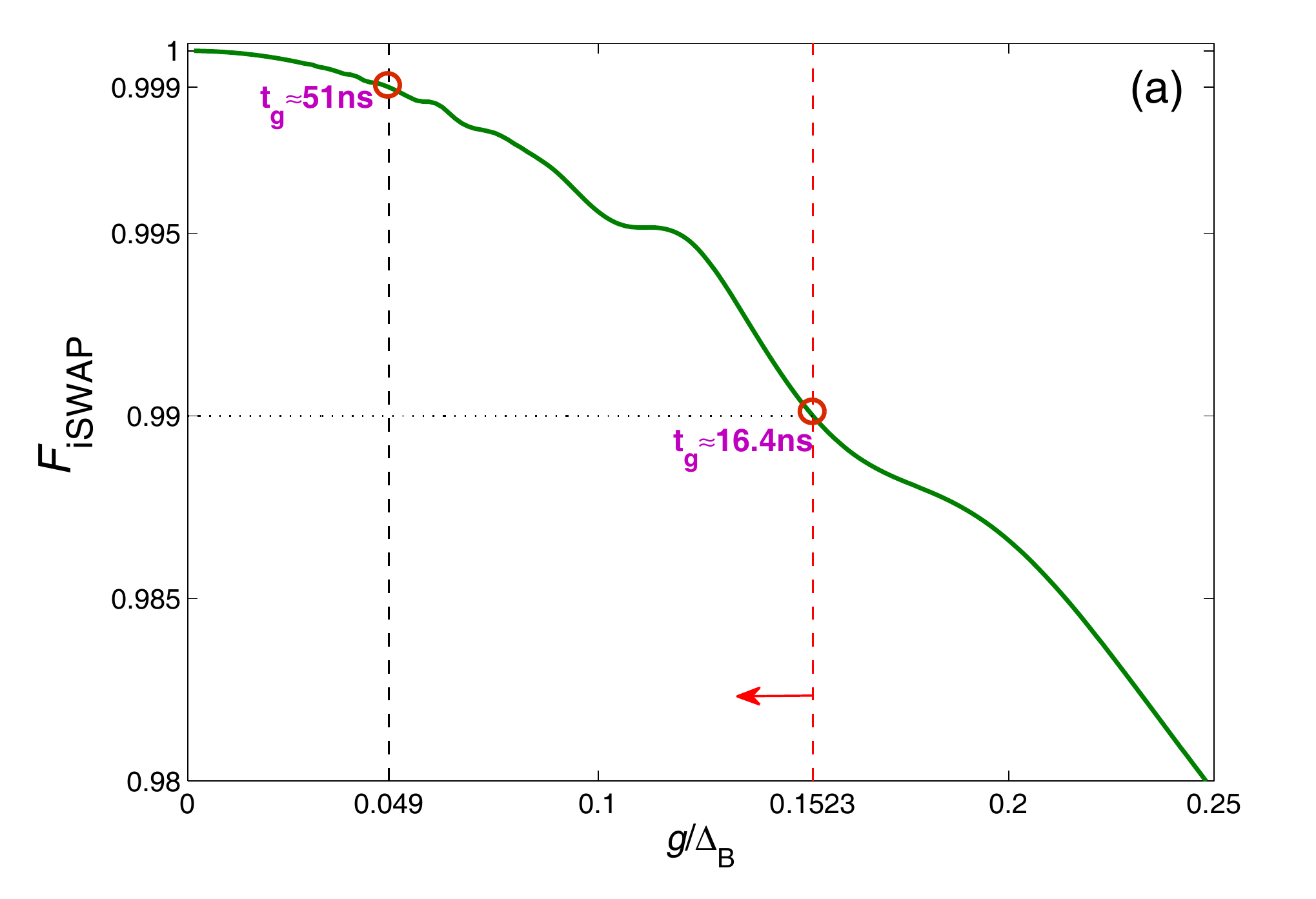}
\includegraphics[width=0.72\textwidth]{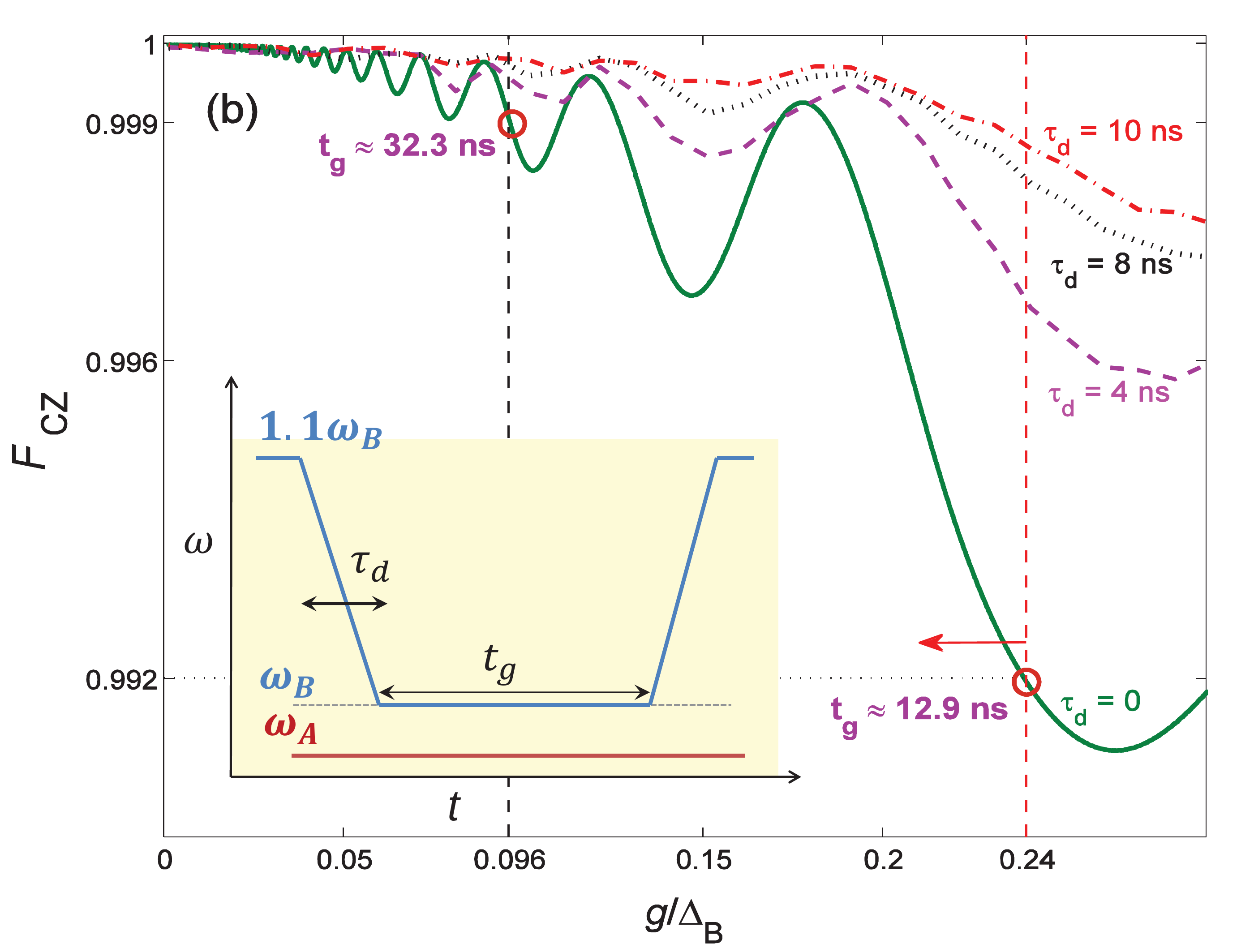}
\caption{(Color online) The fidelities of the two-qubit iSWAP (a)
and CZ (b) gate as functions of $g/\Delta_{B}$ in a circuit with
direct qubit-qubit coupling. Some representative dots are denoted by
the dashed lines and red circles in order to present the
relationship between the gate time $t_{g}$ and fidelity $F$. The red
arrows point out the parameter regime corresponding to two-qubit
gate with high fidelity. In figure (b), the qubit frequencies are
adiabatically adjusted during the gate operation, as shown in the
inset part. The system parameters used here are: (a)
$\omega_{A}/2\pi=5.5$ GHz, $\omega_{B}=\omega_{A}$,
$\Delta_{A}/2\pi=0.15$ GHz, and $\Delta_{B}/2\pi=0.1$ GHz; (b)
$\omega_{A}/2\pi=7.16$ GHz, $\Delta_{A}/2\pi=0.087$ GHz,
$\Delta_{B}/2\pi=0.114$ GHz, and
$\omega_{B}=\omega_{A}+\Delta_{B}$.}
\end{figure}

\begin{figure}[here]
\includegraphics[width=0.52\textwidth]{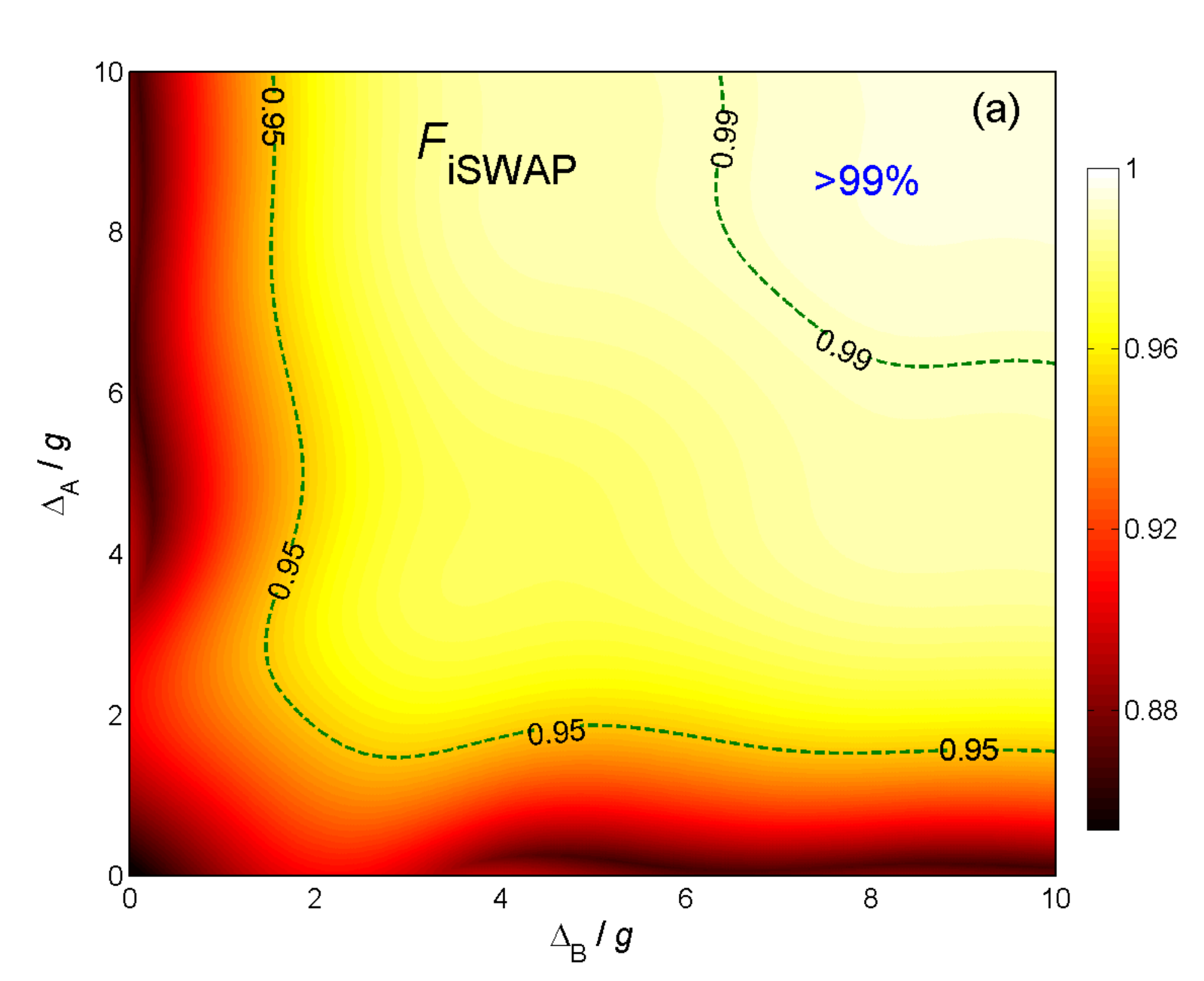}
\includegraphics[width=0.52\textwidth]{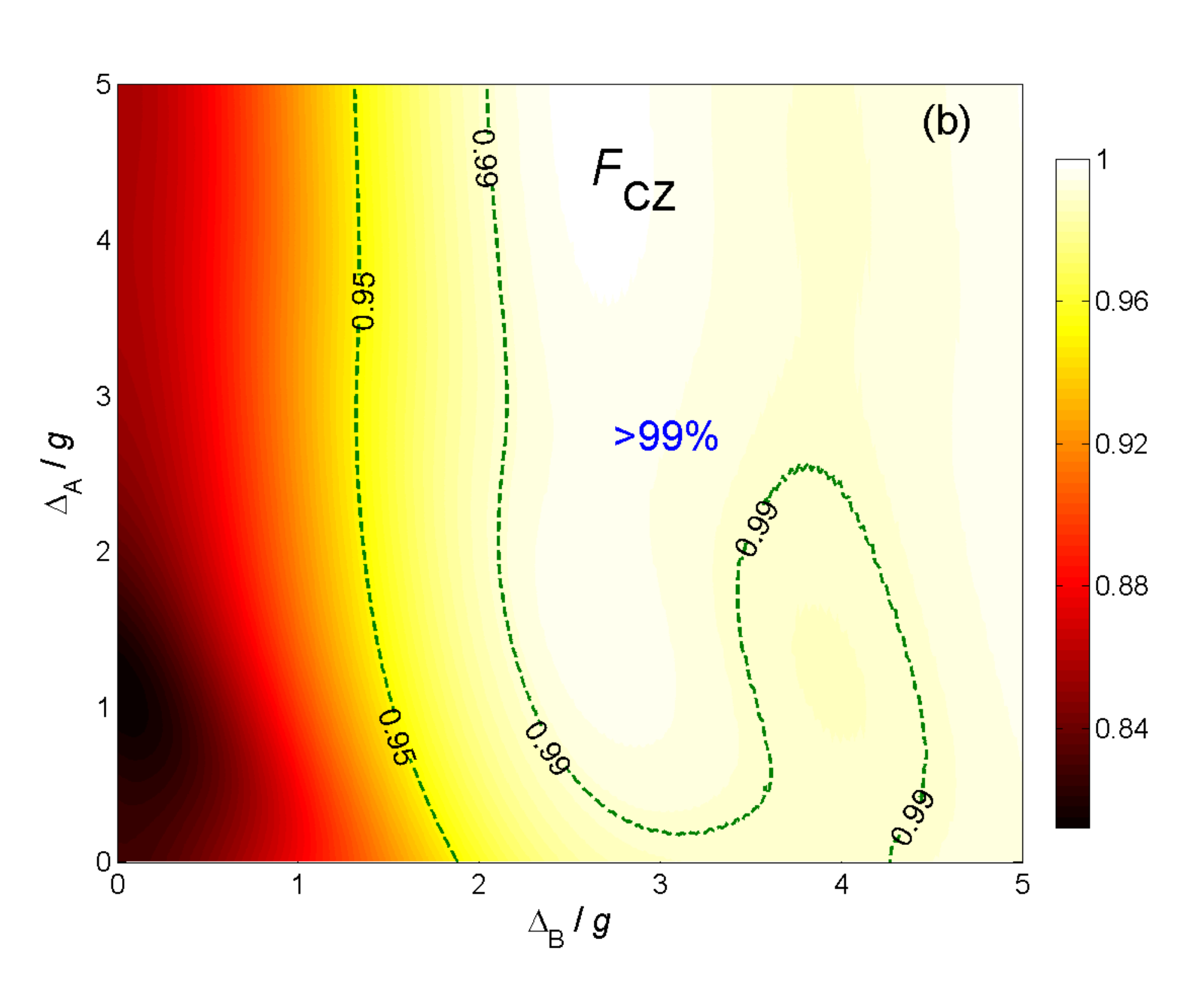}
\caption{(Color online) The fidelities of the two-qubit iSWAP gate
(a) $F_{\rm iSWAP}$ and CZ gate $F_{\rm CZ}$ (b) versus
$\Delta_{A}/g$ and $\Delta_{B}/g$ in a circuit with direct
qubit-qubit coupling. The dashed lines correspond to the parameter
regime for implementing a two-qubit gate with fidelities 95$\%$ and
99$\%$. The system parameters are the same as in Fig.\,3 except for
$g/2\pi=0.2$ GHz.}
\end{figure}

\begin{figure}[here]
\centering
\includegraphics[width=0.7\textwidth]{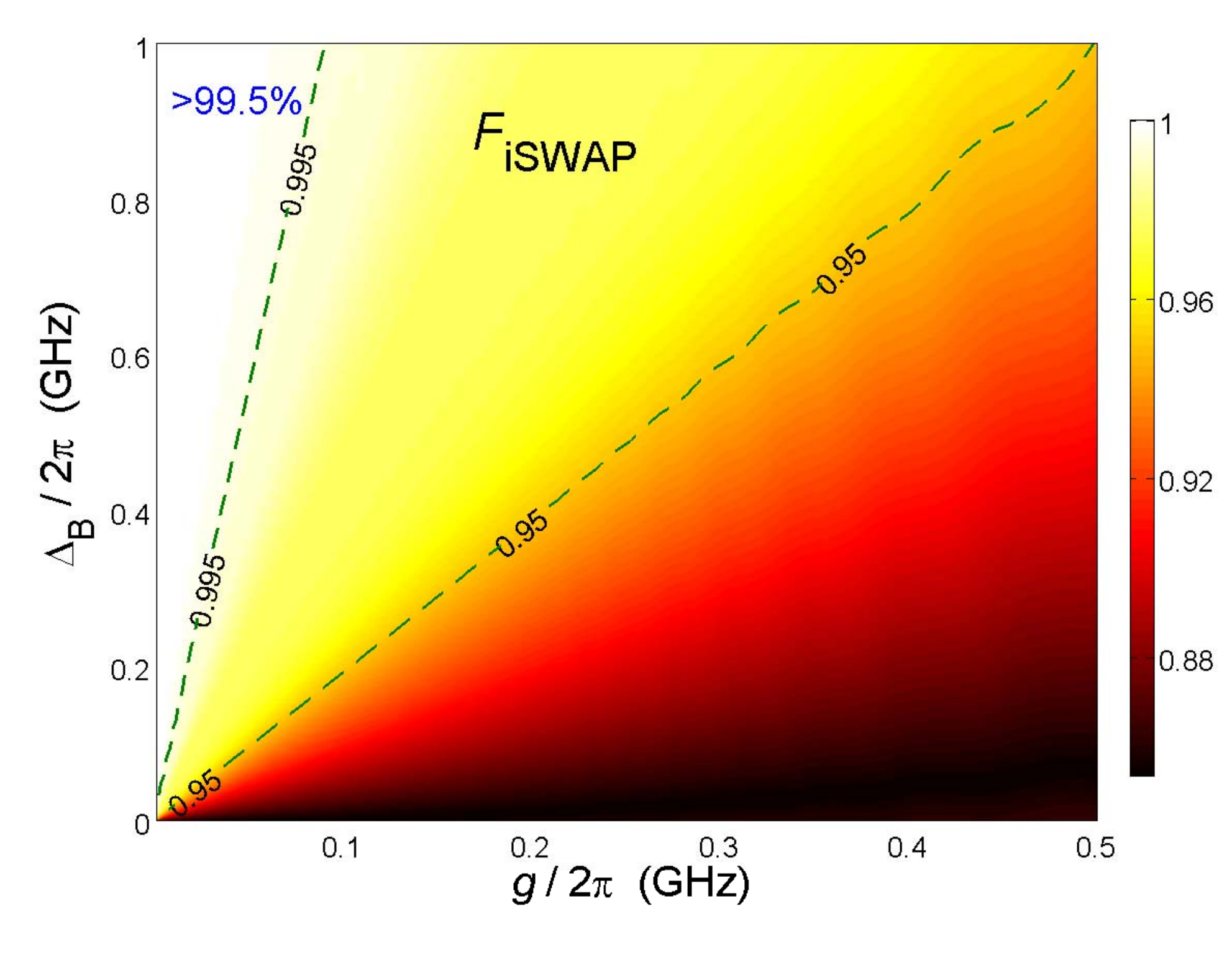}
\caption{(Color online) The fidelities of the two-qubit iSWAP gates
versus $\Delta_{B}$ and $g$ in a circuit with direct qubit-qubit
coupling.  The dashed lines correspond to the parameter regime for
implementing two-qubit gate with fidelities 95$\%$ and 99.5$\%$. The
system parameters are the same as in Fig.\,3 except for
$\Delta_{A}=\Delta_{B}$.}
\end{figure}

In this subsection, based on the original Hamiltonian Eq.\,(1a), we
numerically calculate the influence of the coupling strength $g$ and
anharmonicity $\Delta_{j}$ on the fidelities of the two-qubit iSWAP
and CZ gates (see Figs.\,3-5). Here we consider the two-qubit iSWAP
and CZ gates implemented in experiments \cite{13}. In Figs.\,3(a)
and (b), we plot the fidelities of the two-qubit iSWAP gate ($F_{\rm
iSWAP}$) and the CZ gate ($F_{\rm CZ}$) as functions of
$g/\Delta_{B}$ in a circuit with direct qubit-qubit coupling, where
we consider each SC qubit to have three levels (same approximation
will be used in Figs.\,4 and 5). From Fig.\,3(a) and the (green)
solid line in Fig.\,3(b), it can be seen that the fidelities of
these gates decrease with increasing $g/\Delta_{B}$, and the present
numerical results can help identify the safe parameter regime for
realizing two-qubit gates with high fidelities. As shown in
Fig.~3(a), if we want to implement the two-qubit iSWAP (CZ) gate
with fidelity higher than 99$\%$ (99.2$\%$), the safe parameter
regime is $g/\Delta_{B}<0.152$ ($g/\Delta_{B}<0.24$). In other
words, based on the relationship $gt_{g}=\pi/2$ for the iSWAP gate
and $\sqrt{2}gt_{g}=\pi$ for the CZ gate, the present numerical
results can also identify the time limit for implementing two-qubit
gates with high fidelity. For example, here the shortest gate time
is $t_{g}\approx16.4$ ns ($t_{g}\approx12.9$ ns) for implementing a
two-qubit iSWAP (CZ) gate with fidelity higher than 99$\%$
(99.2$\%$).

The (green) solid line in Fig.\,3(b) shows small oscillations in the
fidelity of the two-qubit CZ gate. This result is due to the
frequency mismatch between the undesired transitions and the
resonant transition [see Fig.~2(b)], and it demonstrated that the
fluctuations of the system parameters will influence the
implementation of two-qubit gates. Based on the idea of
adiabatically eliminating undesired transitions, these oscillations
can be reduced by slowly adjusting the frequencies of the qubits
during the gate operation. As shown in the inset of Fig.\,3(b), the
frequency of qubit B starts at 1.1$\omega_{B}$, is first ramped down
to $\omega_{B}$ in $\tau_{d}$, then ramped up to 1.1$\omega_{B}$
after an interaction time $t_{g}$ $(\sqrt{2}gt_{g}=\pi)$. During the
full gate operation time ($2\tau_{d}$+$t_{g}$), the frequency of
qubit A is fixed. Using such pulses, we numerically calculate the
fidelities of the two-qubit CZ gate for different values of
$\tau_{d}$ and present the results in Fig.\,3(b) [See dashed, dotted
and dot-dashed lines in Fig.\,3(b)]. It can be seen that the
oscillations of the fidelity can be eliminated by adiabatically
adjusting the qubit frequencies during the gate operation. This
numerical result provides a method to reduce the influence of
parameter fluctuations on the implementation of two-qubit gates.
\begin{figure}[here]
\includegraphics[width=0.51\textwidth]{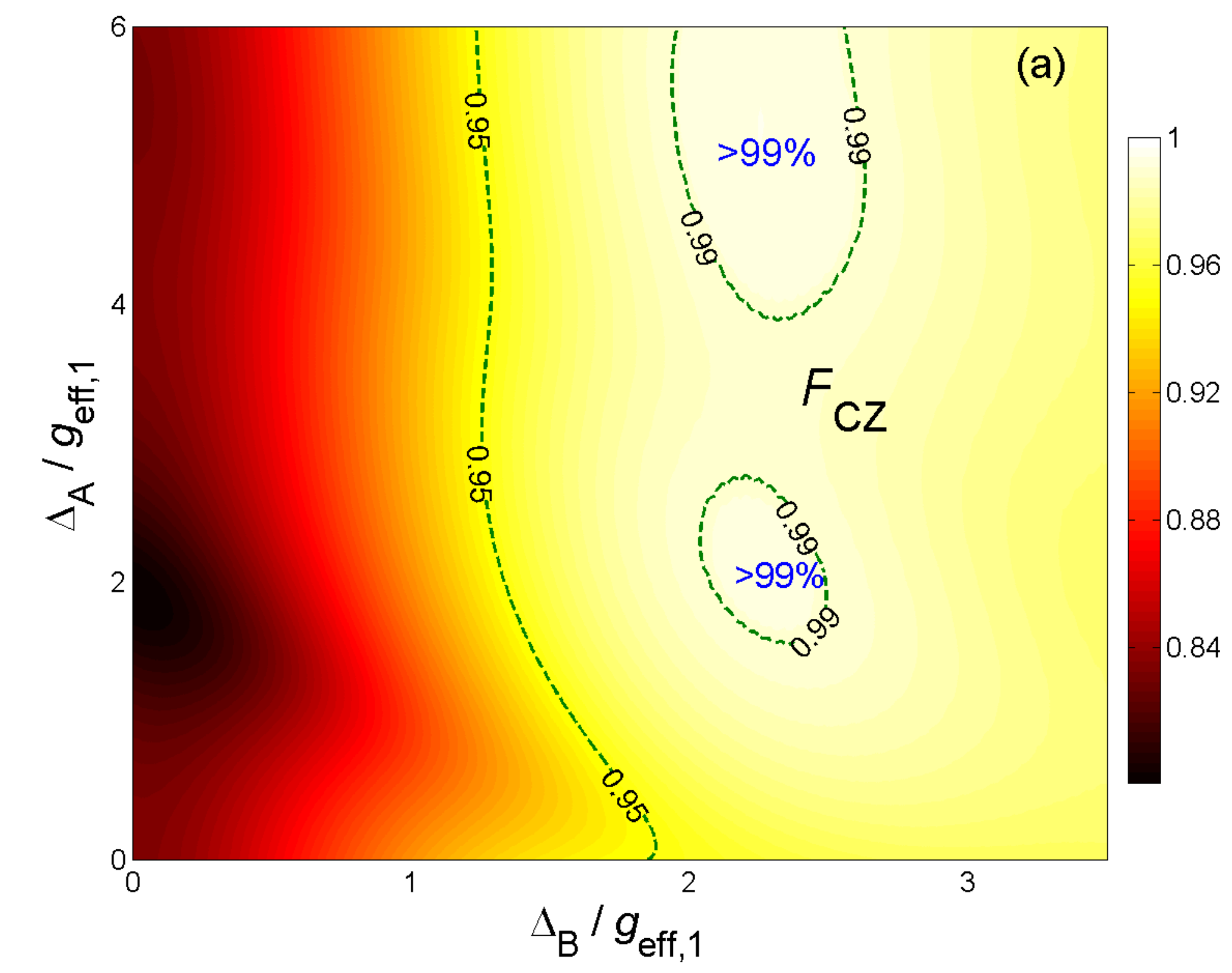}
\includegraphics[width=0.52\textwidth]{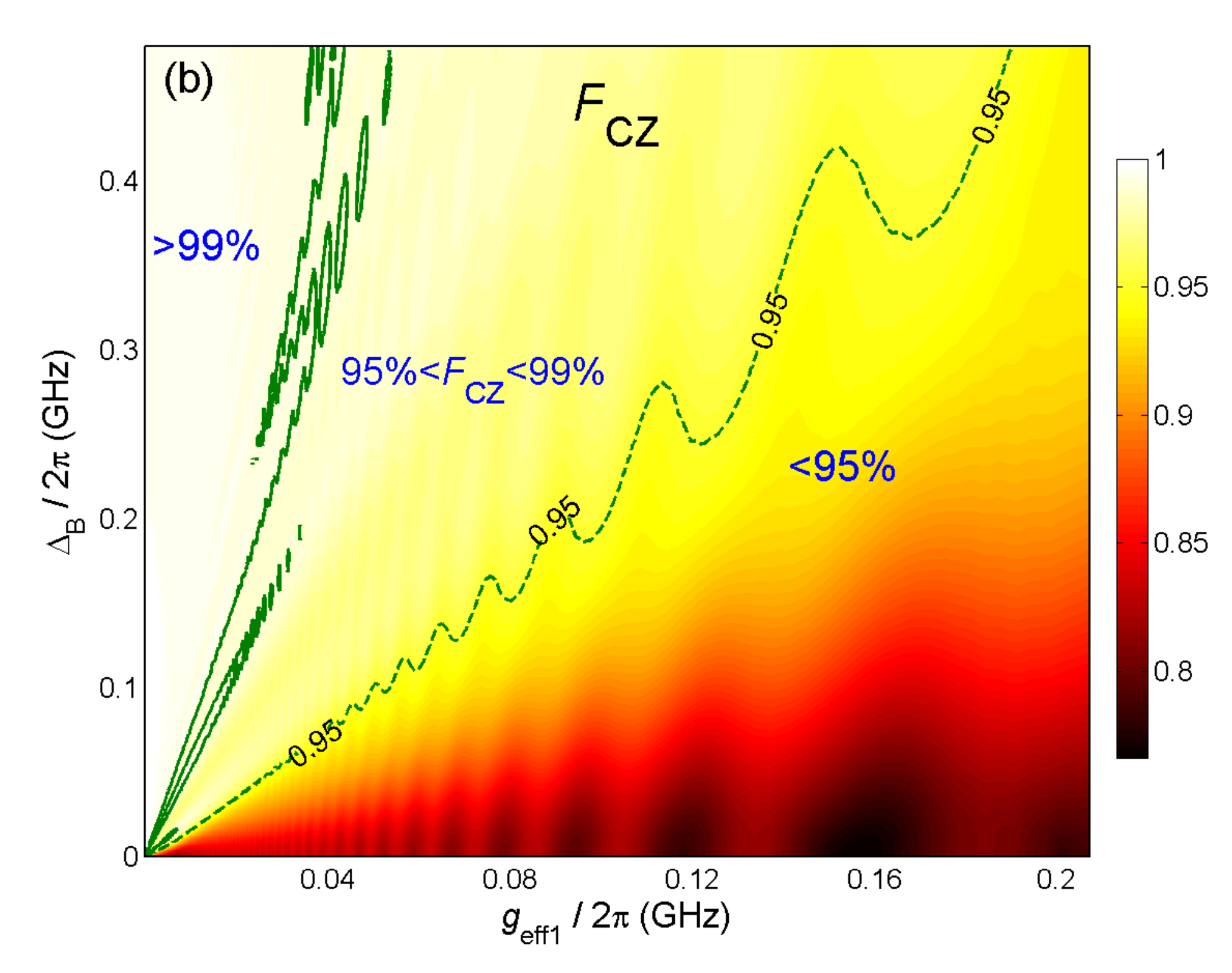}
\caption{(Color online) The fidelity of the two-qubit CZ gate versus
$\Delta_{A}/g_{{\rm eff},1}$, $\Delta_{B}/g_{{\rm eff},1}$ (a) and
versus $g_{{\rm eff},1}$, $\Delta_{B}$ (b) in the system with
indirect qubit-qubit coupling. The dashed lines correspond to the
parameter regime for implementing two-qubit gate with fidelities
95$\%$ and 99$\%$. The basal system parameters are:
$\omega_{c}/2\pi=6.9$ GHz, $\omega_{A}/2\pi=8.2$ GHz,
$\omega_{B}=\omega_{A}+\Delta_{B}$,
$\delta_{j}=\omega_{j}-\omega_{c}$ $(j=A,B)$; And  $G=0.2$ GHz for
panel (a), $\Delta_{A}/2\pi=\Delta_{B}/2\pi$ GHz for panel (b).}
\end{figure}

In order to show the influence of $\Delta_{A}$ and $\Delta_{B}$ on
the two-qubit gates, we plot the fidelities of the two-qubit iSWAP
and CZ gates as functions of $\Delta_{A}/g$ and $\Delta_{B}/g$ in
Fig.\,4. It is easily seen from Fig.\,4(a) that the anharmonicities
$\Delta_{A}$ and $\Delta_{B}$ have equal effects on the two-qubit
iSWAP gate, i.e., the larger the anharmonicities $\Delta_{j}$
$(j=A,B)$ are, the higher the fidelity. This symmetric property
disappears in the two-qubit CZ gate due to the asymmetry in the
condition on the parameters, $\omega_{B}=\omega_{A}+\Delta_{B}$ [see
Fig.\,4(b)]. In other words, the influence of the anharmonicity
$\Delta_{A}$ on the two-qubit CZ gate can be neglected when
$\omega_{B}=\omega_{A}+\Delta_{B}$ is chosen. In addition, the
dashed lines in Fig.~4 indicate the safe regime of $\Delta_{j}/g$
$(j=A,B)$ for implementing two-qubit iSWAP and CZ gates with
fidelity higher than $99\%$.

In Figs.\,3 and 4, either the anharmonicity $\Delta_{j}$ or the
coupling strength $g$ have been set to a fixed value. A natural
question is whether the conclusions obtained from Figs.\,3 and 4 are
universal. In other words, will the properties of Figs.\,3 and 4
change much when either $\Delta_{j}$ or $g$ is changed? Thus, we now
present in Fig.\,5 three-dimensional (3D) plots of the dependence of
$F_{\rm iSWAP}$ on $g$ and $\Delta_{B}$. It is shown that the
fidelity of two-qubit gates are approximately determined by the
ratio of the qubit-qubit coupling strength $g$ to the anharmonicity
$\Delta_{j}$ of the SC qubits. As a result, the conclusion obtained
from Fig.\,3(a) [or Fig.\,4(a)] will not be changed when adjusting
$\Delta_{B}$ (or $g$).  A similar property is also obtained from the
two-qubit CZ gate (the corresponding figures are not shown in this
paper because are very similar to Fig.\,5).

\subsection{System with indirect qubit-qubit coupling}
\begin{figure}[here]
\includegraphics[width=0.52\textwidth]{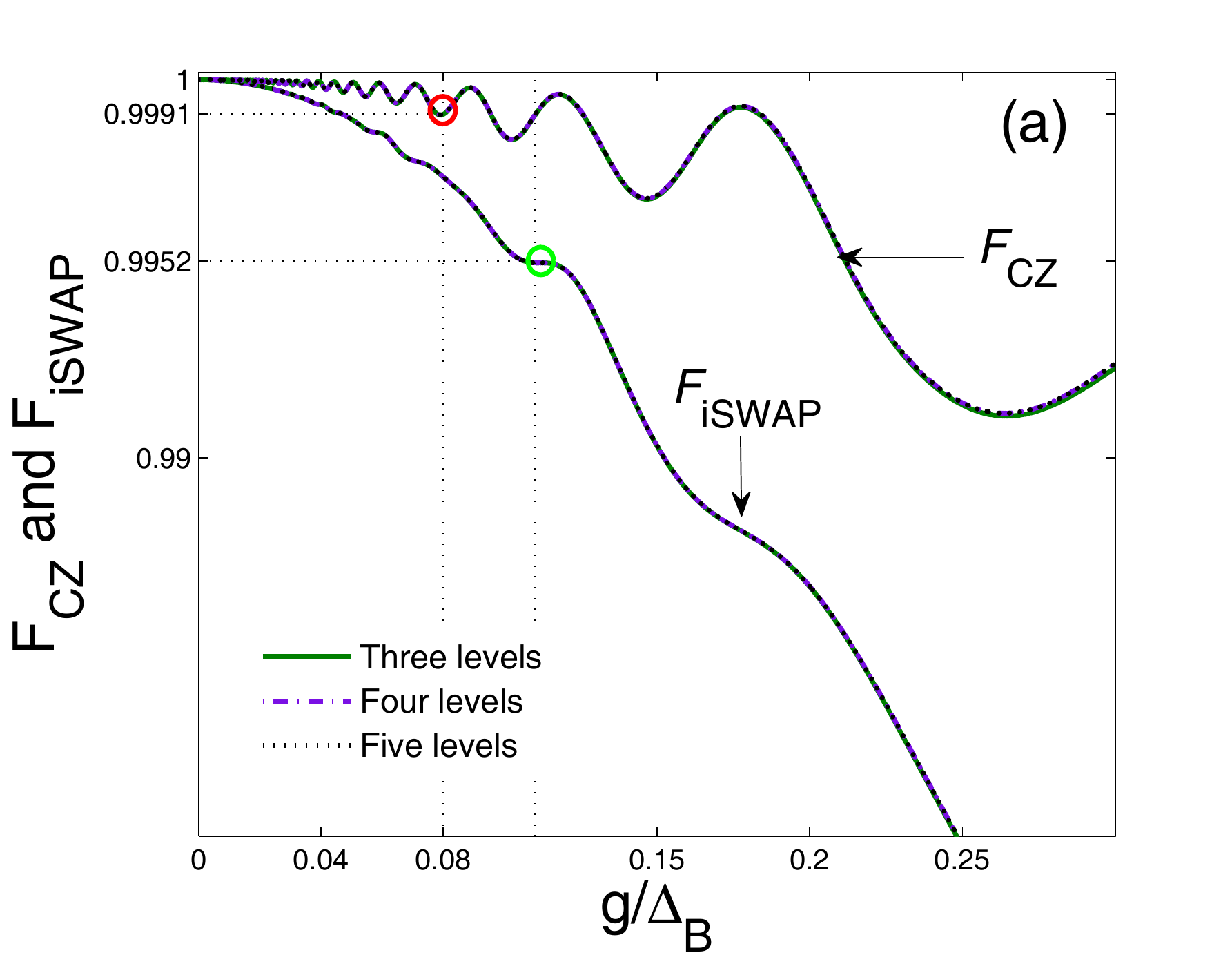} \includegraphics[width=0.52\textwidth]{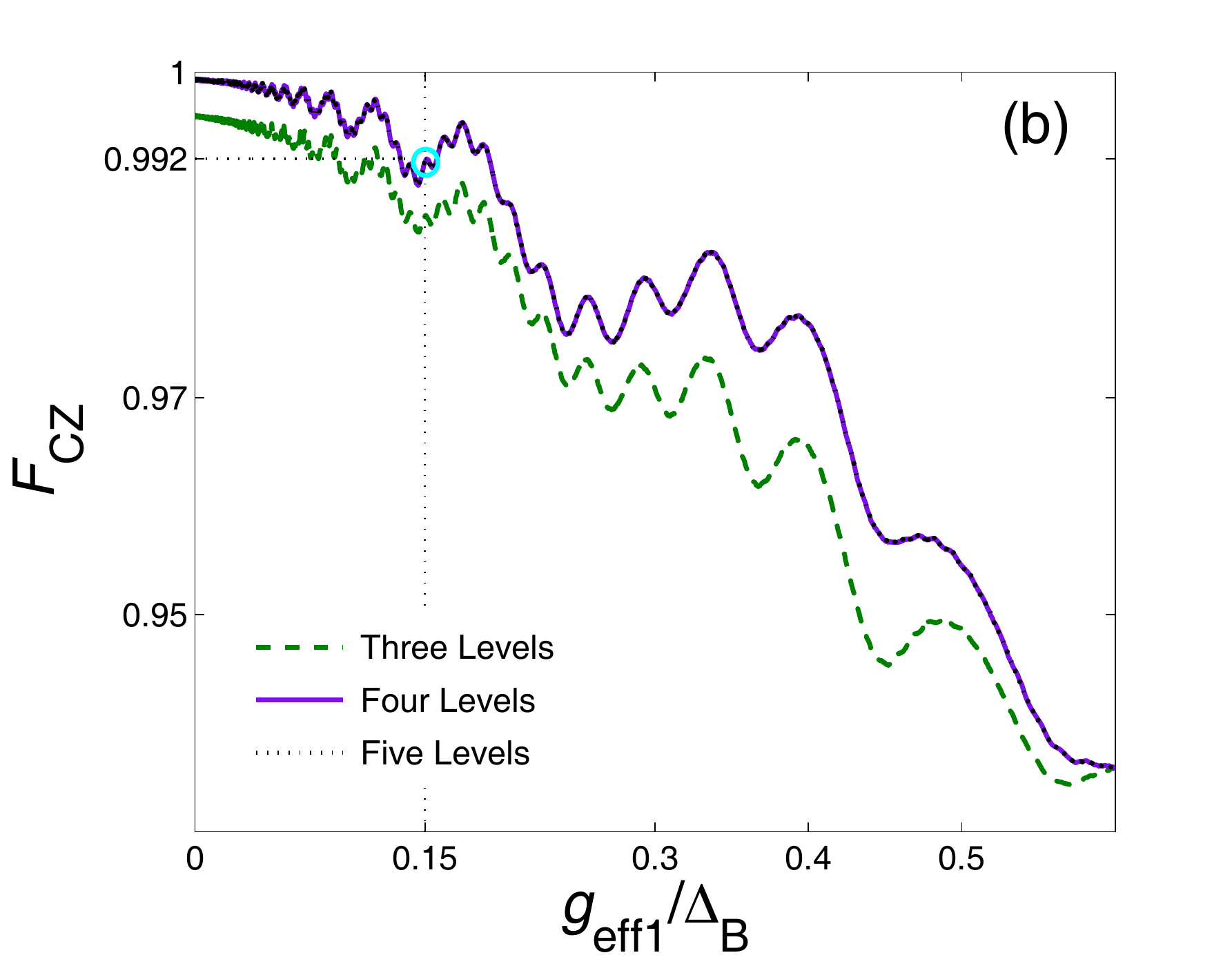}
\caption{(Color online)  The fidelities of the two-qubit gates as a
function of $g/\Delta_{B}$ (a) and $g_{{\rm eff},1}/\Delta_{B}$ (b)
in systems with direct (a) and indirect (b) qubit-qubit coupling,
when the three, four, or five lowest levels are considered for each
qubit. The system parameters are the same as in Fig.\,3 or 6. The
green, red circles in (a) and cyan circle in (b) mark respectively
the experimental parameters regime in Refs.\,\cite{13}, \cite{14},
\cite{15}.}
\end{figure}

In this subsection, based on the Hamiltonian Eq.~(1b), we present
the results of numerical calculations for the dependence of the
fidelity of the two-qubit gates on the effective qubit-qubit
coupling $g_{\rm eff1}$ and anharmonicity $\Delta_{j}$ of SC qubits.
Here the two-qubit CZ gates are realized based on the qubit-cavity
dispersive interaction method \cite{15}, and the parameter
\begin{eqnarray}
g_{{\rm
eff},1}=\frac{G^{2}}{2}\left(\frac{1}{\delta_{B}-\Delta_{B}}+\frac{1}{\delta_{A}}\right)=\frac{G^{2}}{\delta_{A}}\nonumber
\end{eqnarray}
under the condition $\omega_{B}=\omega_{A}+\Delta_{B}$.

In Fig.\,6, we present the 3D plots of the dependence of $F_{\rm
CZ}$ on $\Delta_{A}/g_{{\rm eff},1}$ and $\Delta_{B}/g_{{\rm
eff},1}$ [panel (a)], and $g_{{\rm eff},1}$ and $\Delta_{B}$ [panel
(b)], where we consider the SC qubits to have three levels. Using
dashed lines, we have denoted the parameter regime for implementing
two-qubit CZ gate with fidelities $95\%$ and $99\%$. It is shown
from Figs.\,6(a) and (b) that high-fidelity areas correspond to the
weak-coupling regime $g_{{\rm eff},1}/\Delta_{j}\ll1$ $(j=A, B)$,
while low fidelity corresponds to the strong-coupling regime, where
$g_{{\rm eff},1}$ is comparable to or larger than $\Delta_{j}$. This
property is similar as that in the system with direct qubit-qubit
coupling. The present numerical results can be used to identify the
safe parameter regime for implementing the two-qubit CZ gate with
high fidelity in the circuit with indirect qubit-qubit coupling.

\subsection{Going beyond the three-level approximation}
Until now,  three-level-system approximation for qubits has been
used in the above numerical calculations. It is then natural to ask
the following question: will our conclusions, obtained from the
above numerical results, still be valid for qubits with $N$ ($N>$3)
levels? To explore this, in Fig.~7, we plot the fidelities of the
two-qubit iSWAP and CZ gates as functions of $g/\Delta_{B}$ (or
$g_{{\rm eff},1}/\Delta_{B}$) in the system with direct (or
indirect) qubit-qubit coupling when each qubit has three, four or
five levels. It can be seen from Fig.\,7 that there is not much
difference between the numerical results based on the three-, four-
and five-level approximations for the qubits. So, our conclusions
obtained from the above numerical calculations are still valid for
$N$-level (with $N>$3) SC qubits.

\subsection{Limits on the gate fidelities of recent experiments imposed by weak anharmonicity}
In order to serve as a guide for future experiments, we compare our
numerical results with corresponding experiments and show the
limited fidelity of two-qubit gate based on SC qubits with weak
anharmonicity. Based on the experimental parameters
($\omega_{A}/2\pi$, $\omega_{B}/2\pi$, $\Delta_{A}/2\pi$,
$\Delta_{B}/2\pi$, $g/2\pi$) equal to (5.5, 5.5, 0.15, 0.1, 0.011)
GHz and (7.16, 7.274, 0.087, 0.114, 0.0091) GHz, two-qubit iSWAP
\cite{13} and CZ \cite{14} gates with fidelities 63\% and 70\% were
implemented in the circuit with direct qubit-qubit coupling. In the
circuit with indirect qubit-qubit coupling, a two-qubit gate
\cite{15} with fidelity 85\% was realized with system parameters
($\omega_{c}/2\pi$, $\omega_{A}/2\pi$, $\omega_{B}/2\pi$,
$\Delta_{A}/2\pi$, $\Delta_{B}/2\pi$, $G_{A}/2\pi=G_{B}/2\pi$) equal
to (6.9, 8.2, 8.45, 0.2, 0.25, 0.199) GHz. Corresponding to the
above experimental parameters, in Fig.\,7 we indicate the ideal
fidelity (see the green, red and magenta circles) based on our
theoretical calculations. From the comparison between experiments
and our numerical calculations, we show that two-qubit gates with
fidelities 99.52\%, 99.91\%, and 99.2\% can be realized, in
principle, if the influence of decoherence can be eliminated.
Recently, the effects of decoherence on quantum gates and possible
optimization routes were also studied in Ref.\,\cite{34}.

\section{Conclusion}
We have studied the performance of two-qubit gates in a system of
two coupled SC qubits under the condition that the coupling strength
is comparable to or larger than the anharmonicity of the qubits.
First of all, by using the three-level approximation for the qubits,
we analyzed and numerically calculated the dependence of the
two-qubit gate fidelity on the qubit-qubit coupling strength and the
anharmonicity of the qubits. Based on extensive numerical results,
the safe parameter regime was identified for experimentally
implementing two-qubit gates with high fidelity. Secondly, we
numerically calculated the fidelity of the two-qubit gates in the
case of four- and five-level approximations for the qubits, and
demonstrated the validity of our numerical results for $N$-level
qubits with $N>3$. Our results can serve as a guide for future
experiments based on SC qubits.

\ack We would like to thank E. Solano for useful discussions. This
work was partially supported by ARO grant No.~0726909, JSPS-RFBR
(No.~09-02- 92114), Grant-in-Aid for Scientific Research (S), MEXT
Kakenhi on Quantum Cybernetics, the JSPS via its FIRST program. XYL
was supported by the National Natural Science Foundation of China
(Grant No. 11005057).

\end{document}